\documentclass{aa}  

\usepackage{graphicx}
\usepackage{txfonts}

\begin{document}

   \title{The spin rates and spin evolution of the O components in WR+O binaries}

   \author{D. Vanbeveren\inst{1} \and N. Mennekens\inst{1} \and M. M. Shara\inst{2} \and A. F. J. Moffat\inst{3}}
 
   \institute{Astronomy and Astrophysics Research Group, Vrije Universiteit Brussel, Pleinlaan 2, 1050 Brussels, Belgium\\
    \email {dvbevere@vub.be}
    \and Department of Astrophysics, American Museum of Natural History, Central Park West at 79th Street, New York, NY 10024, USA
		\and Departement de Physique, Universit$\acute{e}$ de Montr$\acute{e}$al, CP 6128 Succ. C-V, Montr$\acute{e}$al, QC H3C 3J7, Canada
    }

   \date{Received }

  \abstract
   {Despite 50 years of extensive binary research we have to conclude that the Roche lobe overflow/mass transfer process that governs close binary evolution is still poorly understood.}
   {It is the scope of the present paper to lift a tip of the veil by studying the spin-up and spin-down processes of the O-type components of WR+O binaries.}
   {We critically analyze the available observational data of rotation speeds of the O-type components in WR+O binaries. By combining a binary evolutionary code and a formalism that describes the effects of tides in massive stars with an envelope in radiative equilibrium, we compute the corresponding rotational velocities during the Roche lobe overflow of the progenitor binaries.}
   {In all the studied WR+O binaries, we find that the O-type stars were affected by accretion of matter during the RLOF of the progenitor. This means that common envelope evolution which excludes any accretion onto the secondary O-star, has not played an important role to explain the WR+O binaries. Moreover, although it is very likely that the O-type star progenitors were spun-up by the mass transfer, many ended the RLOF/mass transfer phase with a rotational velocity that is significantly smaller than the critical rotation speed. This may indicate that during the mass transfer phase there is a spin-down process which is of the same order as, although significantly less than that of the spin-up process. We propose a Spruit-Tayler type dynamo spin-down suggested in the past to explain the rotation speeds of the mass gainers in long-period Algols.}
   {}

   \keywords{binaries: close, binaries: rotation, binaries: magnetic fields}
   
    \authorrunning{D. Vanbeveren et al.}  
    
    \titlerunning{The spin-evolution of the O components in WR+O binaries}
    
   \maketitle

\section{Introduction}

About 30-40\% of the Galactic WR stars have a visible OB-type companion (Vanbeveren and Conti, 1980; van der Hucht, 2001; Crowther, 2007). Detailed massive single star and massive binary population synthesis\footnote {including a model to account for mergers and a formalism to account for the effects on binary parameters of asymmetrical supernova explosions, a process responsible for separating binaries and thus for making single stars which were born as binary components} makes it then possible to answer the question: `What must be the initial O-type close binary frequency to explain the observed WR+OB binary frequency?'. An answer was presented by Vanbeveren et al. (1998a, b, c) who concluded that the initial massive close binary frequency must be at least 70\%, at that time a result that was hard to accept by the massive star scientific community. Fortunately, this predicted high O-type close binary percentage was observationally confirmed about 1.5 decades later by Sana et al. (2012) and it is now generally accepted that most of the massive stars are born in close binaries.

`Close' means that at some point during the binary evolution Roche lobe overflow (RLOF) will happen. RLOF may be accompanied by mass transfer, may result in the formation of a common envelope (note that during a common envelope phase mass transfer is not expected to happen) and/or may result in the merger of both binary components. The interested reader may find useful information regarding these processes and the uncertainties in extended reviews and references therein (van den Heuvel, 1993; Vanbeveren et al., 1998b, c; Langer, 2012; De Marco \& Izzard, 2017). Massive close binary evolution and the study of the binary processes listed above have been the subject of numerous papers since the sixties but it is fair to state that the RLOF with all its facets is still not fully understood and it therefore requires further theoretical and observational research.

Except for about a quarter of Galactic massive WR stars, which are H-rich main-sequence stars of very high mass, most of the remaining Galactic WR-stars are hydrogen deficient core helium burning stars. Their progenitors have therefore experienced extensive mass loss. In the case of single stars this is possibly due to an LBV-type and/or a red supergiant-type mass loss. When the WR-star is a binary component its progenitor may have lost its hydrogen rich layers by RLOF/common envelope evolution. A comparison between the observations of WR+OB binaries and evolutionary prediction may therefore help to understand the RLOF/common envelope evolution of their progenitors. Earlier attempts can be found in e.g., Vanbeveren et al., 1998 b, c; Petrovic et al., 2005a; Eldridge, 2009; Shenar et al., 2016.

Mass transfer is accompanied by angular momentum transfer and the mass gainer is expected to spin up.  When mass transfer/accretion goes via an accretion-disk Packet (1981) showed that this spin-up is very rapid and that soon after the onset of the RLOF the gainer will rotate at its critical break-up velocity. The possible effects of binary interaction on the rotation rates and the rotation velocity distribution of massive stars was investigated by De Mink et al. (2013). It was concluded that a significant percentage of all rapid rotators (maybe all of them) could have a binary origin. The authors note that a main uncertainty affecting this conclusion is the neglect of magnetic fields generated during mass accretion and stellar mergers (see also section 4).   

To test these theoretical expectations the observed spin-rates of the O components in WR+O binaries may be most illuminating. Shara et al. (2017) measured and analyzed the spin rates of the O stars in eight WR+O binaries using the Southern African Large Telescope (SALT) increasing the Galactic sample size from 3 to 11.  The 5 known WR+O binaries of the SMC were investigated by Shenar et al. (2016). The available data are further discussed in section 2 of the present paper.  In sections 3, 4 and 5, we will address the following two questions:

\begin{itemize}
\item 	When the RLOF is accompanied by mass transfer do massive mass gainers always spin-up until they reach their critical rotation velocity?\\ 
\item In anticipation, we will show that the answer to the foregoing question is `no' and that tidal effects are probably not the (only) cause. The second question then is: Is there a physical spin-down process different from tidal effects which is of the same order as, but still less than the spin-up process during RLOF?
 
\end{itemize}

\section{The WR+O binaries: the spin-rates of the O components}

Let us start from Table 2 of Shara et al. (2017). The table lists the Galactic WR+O binaries (11 systems) for which the rotational velocity of the O-component has been determined\footnote{Shara et al. (2017) determined and discussed the rotation speeds by analyzing two He-lines, He I 4922 and He II 4541; here we consider both.} or for which this rotational velocity was discussed in previous work (WR11 from Baade et al. 1990, WR127 from de la Chevrotiere et al. 2011, and WR139 from Marchenko et al. 1994). The table gives the observed mass functions and inclination angle-ranges from Lamontagne et al. (1996). For the WR binaries WR11 and WR139 the inclination angle is quite accurately known and therefore so are the orbital masses\footnote{Note that North et al. (2007) reconsidered WR11 ($\gamma$-Velorum) and proposed very similar mass estimates as those given in Table 2 of Shara et al., 2017}. For the other systems we have further restricted the inclination angle-ranges using the spectral type of the O-type components given by Crowther (2017) and applying the mass-spectral type/luminosity class relation  proposed by Vanbeveren et al. (1998c) for the O-type component \footnote{A mass-spectral type-luminosity class relation relies on evolutionary tracks and therefore depends on parameters whose values are uncertain to some extent; in appendix A we discuss the relation.}. This yields the mass-range estimates of Table 1. Assuming alignment of orbital and spin vectors (not always obeyed: see e.g. Villar-Sbaffi et al.  2005 \& 2006 for the WR+O binaries CQ Cep \& CX Cep, resp.), this restriction then obviously also restricts the possible values of the equatorial velocity. To illustrate the procedure let us consider WR31. It is a WN4o+O8V binary with Msin$^3$i values 2.7 $M_{\odot}$+ 6.3 $M_{\odot}$. The inclination angle-range i = 40-62$^{\circ}$. The mass-spectral type relation predicts a (generous) 24-34 $M_{\odot}$ mass range for an O8V star; this means that an i-value close to 40$^{\circ}$ is to be preferred and therefore an equatorial velocity close to the maximum value given in Shara et al. (2017) should be retained, i.e. 336/274 km/s.  This exercise then yields a set of most probable WR+O star binary properties that are given in Table 2. 

As was already concluded by Shara et al. (2017) the 11 WR+O binaries in our sample have O-type components that seem to be spun up, i.e. the 11 WR+O binaries all show the signature of mass transfer \footnote{Although WR47 looks like an outlier our conclusions related to this binary also rely on the assumption that the WN6o component is hydrogen deficient, post-RLOF and core helium burning}. Since it is expected that a common envelope process is not accompanied by mass transfer we are inclined to conclude that {\it common envelope evolution has not played a significant role in the formation of our WR+O sample}. This also means that the WR stars in our sample of WR+O binaries are not formed by a stellar wind mass loss process only because stellar wind mass loss is not expected to be accompanied by significant mass accretion either. Note that if the 11 WR+O binaries in our sample are representative for the whole Galactic WR+O population, the foregoing conclusions may apply for this whole population.
One can think of two reasons why common envelope evolution may be rare among the most massive binaries: 

a. as was shown by Sana et al. (2012) the period distribution of the most massive binaries may be skewed towards values smaller than the typical periods that are needed for a common envelope phase to happen (of the order of years) and 

b. once a massive star reaches the red supergiant (RSG) phase it is subject to very large stellar wind mass loss (Vanbeveren et al., 2007; Meynet et al., 2015); when a star like that is a component of a binary that would evolve through a common envelope phase if the effects of a RSG wind are ignored, the inclusion of the RSG-wind may significantly reduce the effects of the common envelope process; it may even completely suppress the common envelope phase. Note that most binary population studies (in particular those that tend to predict the merger rates of double BHs by means of isolated close binary evolution) do not account properly for the effect of RSG stellar winds and this makes the predictions very uncertain.

\section{WR+O binaries: the spin-rates of the RLOF progenitors of the O-components}

Some of the WR+O binaries in our sample are very close and they may have been subject to tidal effects which would slow the rotation of the O-type star down; this means that the rotation of the O-stars may have been faster in the past. To estimate the rotation speed of the progenitors of the O-components we proceed as follows. It is generally known that the overall structure of a post-RLOF mass loser of a case B (AB) binary (given its mass) is largely independent of the details of the progenitor evolution in general, the details of the RLOF in particular. Using the Brussels binary evolutionary code we computed the helium burning evolution of a grid of post-RLOF mass losers\footnote {For a description of our evolutionary code and the basic ingredients, see Vanbeveren et al., 1998b, c and Mennekens and Vanbeveren, 2014.}. These computations reveal that Galactic hydrogen deficient WN stars which are mostly early type WN stars (WNE stars using the nomenclature introduced by Vanbeveren and Conti, 1980) are on average 2.10$^{5}$ yrs past their progenitor RLOF; for the WC stars this is on average 3.10$^{5}$ yrs. These timescales hold for WR stars of the binaries listed in Table 2. Given a WR+O system with the parameters of Table 2 we interpolate in the grid in order to estimate how the binary looked immediately after RLOF, using the timescales given above and assuming that during these short WR-timescales the structure of the O-component does not change (a very reasonable assumption accounting for the fact that the O-component is a core hydrogen burning star evolving on a timescale that is typically 10-times longer than the WR-timescale). It is then straightforward, using a scenario that treats tidal effects for stars with a radiative envelope (we used the formalism of Zahn, 1977) to calculate backwards the rotation speed of the O star progenitor at the end of the RLOF. The results are shown in Table 2 as well (v$_{rot}$ of the O-type star at the beginning of the WR phase of the binary is of course v$_{rot}$ of the O-type component at the end of the RLOF). Note that (as is well known) tides are effective mainly in the shortest period binaries. Accounting for the fact that the critical rotation velocity of the O-type stars in our sample ranges between 600 km/s and 800 km/s the results lead us to conclude that

\begin{itemize}
\item most of the mass gainers in the progenitors of WR+O binaries are spun up by mass transfer but their rotation velocity at the end of the RLOF may be significantly smaller than the critical value. The latter is a fortiori true when the rotational velocities are considered resulting from the He II lines.
\end{itemize}

The foregoing conclusion also applies for the longest period WR+O binaries where the effects of tides (and their uncertainties) are small.

As a note added in proof, in a recent paper (Shenar et al., 2016) the authors have investigated the 5 known WR+O binaries in the SMC. The O-type stars seem to rotate super synchronously but also here the rotation speed may be far below critical. In particular, the binary AB8 consists of a WO4 component with an O4V companion. The O4V star has an apparent age of $\sim$10$^{6}$ yrs (see also Appendix A) which is similar to the ages of O4V-III stars in the Magellanic Clouds determined by Massey et al. (2000). One concludes that the O4V component is rejuvenated with respect to the age of the WO4 star (3.10$^{6}$ yrs; see Shenar et al., 2016) meaning that mass transfer happened. The rotation speed of the O-type star lies in the range 130-230 km/s. Since the binary period = 16.6 days the effects of tidal interaction are small so that the present rotation speed of the O-type star almost equals the speed at the end of the RLOF of the progenitor. Compared to the critical rotation speed (which is of the order of 700 km/s using the system parameters tabulated by Shenar et al., 2016) we conclude that also here it is very likely that the mass gainer has been spun-up by mass transfer but at the end of RLOF its super-synchronous rotation velocity was considerably smaller than the critical speed. 

\section{The WR+O binaries: why the progenitors of the O-components spin up during RLOF but may not reach the critical speed}

The conclusions of the previous section lead us to suspect that during the RLOF of the progenitors of the WR+O binaries there must be an efficient spin-down process, otherwise the O components would all be observed to rotating at or near critical speed. Since our WR+O sample reveals that all O-type stars rotate super-synchronously the magnitude of the spin-down process should be of the same order as the spin-up but it should not completely suppress the spin-up. Note that this also applies for the longer period systems where the effects of tides (and their uncertainties) are expected to be small. Let us first remind the interested reader of a previous study (Dervisoglu, Tout and Ibanoglu, 2010) where the authors investigated the spins of the mass gainers of Algol binaries. In long-period Algols (P $>$ 5 days) mass transfer happens through a disc, implying a rapid spinning-up of the mass gainer. According to the analysis of Packet (1981) the rotation speed is expected to become critical soon after the onset of the mass accretion.  However, the observations of longer period Algols reveal that the mass gainers have spins typically between 10 and 40\% of the critical rate and therefore an efficient spin-down mechanism for the gainers is required.  The authors argue that radiative tides are too weak to account for sufficient spin-down and they propose `{\it the generation of magnetic fields in the radiative atmospheres of a differentially rotating star (the Spruit-Tayler type dynamo, Spruit, 2002 and Tayler, 1973) and the possibility of angular momentum loss driven by strong magnetic stellar winds (the model of Tout and Pringle, 1992)}'. Dervisoglu et al. (2010) presented a possible mathematical formalism for this process and concluded that the gainers in Algols with orbital periods longer than $\sim$5 days rotate below breakup, when a small amount of mass, ($\sim$10\% of the transferred matter) is lost by the gainer which has a rotationally induced magnetic field of the order of one or a few kG.

Since the Algol-situation shows some resemblance with the WR+O binary situation (mass transfer caused the spin-up of the mass gainer but the speed of the gainer at the end of RLOF is significantly below critical), we were tempted to test if the same process could work for the massive binaries as well. We follow the formalism described by Dervisoglu et al. and compare the following two processes.

When during RLOF mass is transferred at a rate $\dot{M}_{acc}$  from a disc to the star (with mass M and radius R) the rate of angular momentum transferred from the disc to the outer layers of the star is

\begin{equation}
\frac{dJ_{acc}}{dt} = \dot{M}_{acc}\sqrt{GMR}
\end{equation}

\noindent (G is the gravitational constant). 	  

	The angular momentum lost from a magnetic star due to a wind that is corotating up to the Alfvén surface (with radius $R_{A}$) is

\begin{equation}
\frac{dJ_{w}}{dt} = \dot{M}_{w}{R_{A}}^{2}\Omega
\end{equation}

\noindent with $\dot{M_{w}}$ the wind mass loss rate and $\Omega$ the angular velocity of the outer layers of the star. If it is assumed that the generated magnetic field (with magnetic flux density at the stellar surface $B_s$ expressed in Gauss) is a dipole field, a straightforward calculation gives (using cgs units)

\begin{equation}
\frac{dJ_{w}}{dt} = -(-\dot{M}_{w})^{3/7}{B_{s}}^{8/7}(2GM)^{-2/7}R^{24/7}\Omega
\end{equation}

	Dervisoglu et al. (2010) argued that a fully efficient Spruit-Tayler dynamo can generate a dipole field of the order of one or a few kG and it can support a wind with a rate up to 0.01 $M_{\odot}$/yr. This rate is comparable to the mass loss rate of the mass loser during the RLOF in intermediate-mass close binaries (forming the Algols) and in massive close binaries (forming the WR+O systems). If $\dot{M}_{RLOF}$ is the mass loss rate of the mass loser during its RLOF it is useful to write $\dot{M}_{acc}=-\beta\dot{M}_{RLOF}$ 
and 
$\dot{M}_{w}=(1-\beta)\dot{M}_{RLOF}$

Note that the Spruit-Tayler dynamo operates as long as the star is differentially rotating, i.e. as long as mass transfer happens. If mass transfer stops the star is expected to regain a state of uniform rotation and the magnetic field may vanish. Post-RLOF binaries or case A binaries \footnote{Case A binaries are very short period binaries where RLOF starts while the mass loser is a core hydrogen burning star; case A RLOF starts with a rapid mass transfer phase on the thermal timescale of the loser, followed by a slow phase acting on the nuclear timescale of the loser, i.e. the slow phase lasts almost the entire remaining core hydrogen burning phase.} during the slow phase of RLOF may therefore not show the remains of such fields. We will come back to this later. 

To illustrate whether or not the magnetically induced spin-down (formula 3) can compensate for the spin-up (formula 1) let us consider the evolution of a 30 $M_{\odot}$ + 20 $M_{\odot}$ zero age main sequence binary with a period assuring a case B type RLOF (we use the computations of de Loore and Vanbeveren, 1995). The computations reveal that during the rapid (resp. slow) phase of the RLOF the mass loss rate of the loser ($\dot{M}_{RLOF}$) typically equals 5.10$^{-3}$ $M_{\odot}$/yr (resp. 10$^{-3}$ $M_{\odot}$/yr). Tables 3 and 4 show the ratio  dJ+/dJ- = $\vert$expression (1)/expression (3)$\vert$ for the two $\dot{M}_{RLOF}$ values, for different values of the magnetic field strength and different values of $\beta$. Note that a ratio smaller than 1 corresponds to a situation where rapid spin-up is compensated. The results illustrate the following conclusion

\begin{itemize}
\item
When during the mass transfer phase of a massive binary the Spruit-Taylor dynamo can generate a magnetic field of the order of a few kG then rapid spin-up of the mass gainer can be slowed down at the expense of a moderate mass loss from the binary.
\end{itemize}

A more detailed calculation of this effect would require the implementation of rotation and magnetic fields in the stellar structure equations (as was done by Petrovic et al., 2005b, see also Potter et al., 2012) but this is beyond the scope of the present paper.

\section{Observational signature of magnetic fields in the O-type components of massive close binaries}

Plaskett's star (HD 47129) is a massive binary (O8I+O7.5III) with a 14.4-day orbital period (Linder et al., 2008; Mahy et al., 2011). When accounting for the inclination angle range i = 69.3$^\circ$-72.7$^\circ$ proposed by Bagnuolo et al. (1992) both components have a mass between 50 $M_{\odot}$ and 60 $M_{\odot}$.  The secondary is a fast rotator (v$_{rot}$sini $\sim$300 km/s) and it was argued by Linder et al. (2008) that the secondary may rather be an O7V star showing an apparent O7.5III spectrum because of its rapid rotation. Compared to the primary star the secondary is then rejuvenated indicating that mass transfer has happened and that the primary is on its way to become a WN star. It is tempting to conclude that the secondary was spun up by mass transfer but since the critical rotation speed of an O7V star is about 800 km/s, the mass gainer did not reach this critical limit (note that tides in a binary with a 14.4-day period are not expected to be very effective so that the presently observed rotation speed almost equals the speed at the end of the previous RLOF), and all this is much like in many WR+O binaries in our sample. Most interestingly, a strong magnetic field was detected in the secondary (B $>$ 2.85 kG,  Grunhut et al., 2013) and this makes Plaskett's star a strong candidate for the Spruit-Tayler dynamo scenario for the generation of kG-like magnetic fields in RLOF/mass transferring binaries (see also Langer, 2014). The results of Tables 3 and 4 then illustrate that this magnetic field may be responsible for sufficient spin-down explaining why the mass gainer did not reach the critical speed.

Naze et al. (2017) observed a sample of 15 interacting and post-interaction O-type binaries and, very interestingly, they found no indication of a magnetic field in any of the 15 stars and the authors conclude that binary interactions do not systematically trigger stable, strong magnetic fields in such systems. Most of the binaries in the sample of Naze et al. (2017) are very short period systems which makes them case A binary candidates during the slow phase of mass transfer. During the slow phase in a case A binary, rapid spin-up of the gainer is not expected to happen so that a Spruit-Tayler dynamo may not be operational. The results of Naze et al. support this view. Note that most of the binaries in the sample of Naze et al. (2017) are distinctly different from Plaskett's star which, according to its period, is most likely a case B binary with a mass transfer history that is quite different from that in case A binaries. 

\section{Conclusions}

The rotational velocities of the O-type components in 8 WR+O binaries have been measured by Shara et al (2017) using SALT; 3 measurements of the O-type component of Galactic WR-binaries and at least 1 in the SMC were available in the literature. All 12 rotate super-synchronously and we consider this as strong evidence that they were spun up during the RLOF/mass transfer of the progenitor binary. We conclude that common envelope evolution (meaning no mass transfer and thus no spin-up) has not played an important role in the formation of these 12 binaries, and generalizing, that common envelope evolution does not happen frequently in the massive binaries that produce WR+O systems. We implemented the formalism describing the effects of tides in stars with a radiative envelope in our binary evolutionary code and we computed the rotation speed of the O-type stars at the end of the RLOF of the progenitor binary.  Many of them rotate significantly more slowly than critical. As an explanation we propose a model that has been worked out for long period Algols (period $>$ 5 days) where the mass gainers do not rotate at the critical speed either, i.e., - mass transfer causes the mass gainer to rotate differentially - differential rotation makes the Spruit-Tayler dynamo operational and a magnetic field is generated - when the magnetic field is of the order of a few kG a magnetic wind sets in which the mass loss rate is of the order of the mass transfer rate of the mass loser - the combination of wind and magnetic field spins the star down at a rate which is comparable to the mass accretion induced spin-up. In Plaskett's star the mass gainer has been spun-up but also here the critical speed was not reached. The star has an observed magnetic field B $>$ 2.85 kG and is therefore an observational test bed for the spin-up-spin-down scenario outlined above.

\begin{table*}
\caption{The mass-range of the O-type components in WR+O binaries resulting from the observed spectral type.}
\label{table1}
\centering
\begin{tabular}{c c c}
\hline
System & Sp. Type & Mass estimate of O-star ($M_{\odot}$) \\
\hline
WR21 & WN5o+O4-5 & 37-60 \\
WR30 & WC6+O6-8 & 24-50 \\
WR31 & WN4o+O8V & 24-34 \\
WR42 & WC7+O7V & 30-37 \\
WR47 & WN6o+O5V & 37-70 \\
WR79 & WC7+O5-8 & 24-60 \\
WR97 & WN5b+O7 & $>$30 \\
WR113 & WC8d+O8-9IV & 20-30 \\
WR127 & WN5o+O8.5V & 17-24 \\
\hline
\end{tabular}
\end{table*}

\begin{table*}
\caption{Most probable masses and rotation speed of the O-type component of the 11 WR+O binaries in our sample. Column 7 lists the rotation velocities at the end of the RLOF (= the beginning of the WR phase) of the progenitor binary calculated as explained in the text. The v$_{rot}$ column 6 lists the speed values resulting from the He I/He II line measurements as explained in Shara et al. (2017). A question mark means that this particular measurement is not available.}
\label{table4}
\centering
\begin{tabular}{c c c c c c c}
\hline
System & Sp. Type & P (days) & WR mass & O-mass & v$_{rot}$ (km/s) & v$_{rot}$ (km/s) \\
       &          &          &         &         &                 & begin WR         \\
\hline
WR21 & WN5o+O4-5 & 8.3 & $>$19 & $>$37 & 331/138 & 415/173 \\
WR30 & WC6+O6-8 & 18.8 & 16.4 & 34 & ?/$>$178 & ?/$>$205 \\
WR31 & WN4o+O8V & 4.8 & $>$11 & $>$24 & 336/274 & 493/402 \\
WR42 & WC7+O7V & 7.9 & 16 & 27 & 500-511/170-177 & 620-630/211-218 \\
WR47 & WN6o+O5V & 6.2 & 52 & 60 & ?/$>$88 & ?/$>$190 \\
WR79 & WC7+O5-8 & 8.9 & $>$9.5 & $>$24 & ?/$>$174 & ?/$>$250 \\
WR97 & WN5b+O7 & 12.6 & 18 & 30 & 470/279 & 502/298 \\
WR113 & WC8d+O8-9IV & 29.7 & $>$11 & $>$22 & 310-330/170-181 & 320-340/175-186 \\
WR11 & WC8+O7.5III-V & 78.5 & 9.6 & 30 & 220/? & 232/? \\
WR127 & WN5o+O8.5V & 9.5 & $>$9 & $>$20 & 300-366/? & 350-416/? \\
WR139 & WN5o+O6III-V & 4.2 & 9.4 & 28 & 215/? & 365/? \\
\hline
\end{tabular}
\end{table*}

\begin{table*}
\caption{dJ+/dJ- as function of $\beta$ and $B_s$ for $\dot{M}_{RLOF}$ = 10$^{-3}$ $M_{\odot}$/yr}
\label{table5}
\centering
\begin{tabular}{c c c c c c}
\hline
$\beta$ & $B_s$ = 500 G & $B_s$ = 1 kG & $B_s$ = 2 kG & $B_s$ = 3 kG & $B_s$ = 4 kG \\
\hline
0.9 & 9 & 4.1 & 1.8 & 1.2 & 0.8\\
0.8 & 6 & 2.7 & 1.2 & 0.8 & 0.6\\
0.6 & 3 & 1.5 & 0.7 & 0.4 & 0.3\\
0.4 & 1.9 & 0.8 & 0.4 & 0.2 & 0.2\\
0.2 & 0.8 & 0.4 & 0.2 & 0.1 & 0.1\\
0.1 & 0.4 & 0.2 & 0.1 & 0.05 & 0.04\\
\hline
\end{tabular}
\end{table*}

\begin{table*}
\caption{dJ+/dJ- as function of $\beta$ and $B_s$ for $\dot{M}_{RLOF}$ = 5.10$^{-3}$ $M_{\odot}$/yr}
\label{table4}
\centering
\begin{tabular}{c c c c c c}
\hline
$\beta$ & $B_s$ = 500 G & $B_s$ = 1 kG & $B_s$ = 2 kG & $B_s$ = 3 kG & $B_s$ = 4 kG \\
\hline
0.9 & 22.5 & 10.2 & 4.6 & 2.9 & 2.1\\
0.8 & 15 & 6.7 & 3 & 1.9 & 1.4\\
0.6 & 8.3 & 3.8 & 1.7 & 1.1 & 0.8\\
0.4 & 4.6 & 2.1 & 1 & 0.6 & 0.4\\
0.2 & 2.1 & 0.9 & 0.4 & 0.3 & 0.2\\
0.1 & 1 & 0.4 & 0.2 & 0.1 & 0.05\\
\hline
\end{tabular}
\end{table*}

\appendix
\section{The O-type star calibration}

A calibration that links the spectral-type-luminosity class of O-type stars and the luminosity-effective temperature has been presented by Humphreys \& McElroy (1984). Using a set of massive star evolutionary calculations, it is then straightforward to determine a spectral-type/luminosity class-mass relation. This was done by Vanbeveren et al. (1998c) using the evolutionary tracks of Schaller et al. (1992). Table A1 summarizes the calibration for the O-type companions in our WR binary sample which are either luminosity class V or III (the column labeled old). The Schaller et al. tracks do not account for rotation and they make use of stellar wind mass loss rate formalisms that are outdated. Since it has become clear that O-type stars may be rapidly rotating we repeated the exercise done in Vanbeveren et al. but using the most recent Geneva tracks calculated with the most up-to-date mass-loss rate recipes and with an initial rotation speed of 40\% the critical one. This corresponds to the following velocities on the ZAMS: a 9 $M_{\odot}$ star rotates at 230 km/s, a 20 $M_{\odot}$ at 275 km/s, a 40 $M_{\odot}$ at 315 km/s, a 60 $M_{\odot}$ at 345 km/s and a 120 $M_{\odot}$ at 390 km/s. The models used here were downloaded from the Geneva site $obswww.unige.ch/Recherche/evol/tables\_grids2011/Z014/$. The resulting mass calibration is given in Table A1 as well (the column labeled new). As can be noticed the differences between old and new are very small, too small to affect the overall conclusions of the present paper. Note that these small differences are an indication that the calibration of Table A1 is quite robust. 

The O-companions of the WR binaries in our sample often have luminosity class V. Similarly as for the mass calibration, we can assign an age as a function of spectral-type/luminosity class using the two sets of evolutionary tracks discussed above. The resulting calibration is given in Table A2 for OV-stars. As for the mass calibration, the lifetime relation for the OV stars does not depend much on the chosen tracks. The results of Table A1 illustrate that when an OV star is a component of a WR+O binary where the WR star is core helium burning, the OV star is rejuvenated, e.g. mass transfer happened during the previous RLOF.   

\begin{table*}
\caption{The mass/spectral type/luminosity class relation of O-type stars using the evolutionary tracks without rotation of Schaller et al. (1992) (labeled old) and using the evolutionary tracks with rotation from the Geneva data base (labeled new).}
\label{table2}
\centering
\begin{tabular}{c c c}
\hline
Sp. type + Lum. class & Old mass estimate  ($M_{\odot}$) & New mass estimate ($M_{\odot}$) \\
\hline
O4V & 56.1 & 55 \\
O5III & 58.8 & 53 \\
O6V & 37.1 & 36.4 \\
O6III & 49.3 & 42.5 \\
O7.5V & 32.7 & 28 \\
O7.5III & 33.5 & 30 \\
O9V & 23.6 & 22.5 \\
O9III & 26.7 & 25 \\
\hline
\end{tabular}
\end{table*}

\begin{table*}
\caption{The lifetime (in Myr)/spectral type relation of O-type stars with luminosity class V using the evolutionary tracks without rotation of Schaller et al. (1992) (labeled old) and using the evolutionary tracks with rotation from the Geneva data base (labeled new).}
\label{table3}
\centering
\begin{tabular}{c c c}
\hline
Sp. type + Lum. class & Old lifetime  (Myr) & New lifetime (Myr) \\
\hline
O4V & 0.8 & 1 \\
O6V & 1.8 & 1.8 \\
O7.5V & 3.1 & 3.3 \\
O8.5V & 3.7 & 4.0 \\
O9V & 4.1 & 4.5 \\
\hline
\end{tabular}
\end{table*}

\begin{acknowledgements}

We thank the unknown referee for her/his very valuable remarks and suggestions.
      
\end{acknowledgements}


\begin{thebibliography}{}

  \bibitem[Baade(1990)]{Baade1983} Baade, D., Schmutz, W. \& van Kerkwijk, M., 190, A\&A, 240, 105
  
    \bibitem[Bagnuolo(1992)]{Bagnuolo1992} Bagnuolo, W.H., Jr., Gies, D.R. \& Wiggs, M.S., 1992, ApJ, 385, 708
    \bibitem[Crowther(2007)]{Crowther2007} Crowther, P., 2007, ARA\&A, 45, 177

    \bibitem[Crowther(2017)]{Crowther2015} Crowther, P., 2017, Galactic Wolf-Rayet Catalogue, Univ. of Sheffield

	
   \bibitem[Chevro(2011)]{Chevro2011} De la Chevrotiere, A., Moffat, A.F.J. \& Chene, A.-N., 2011, MNRAS, 411, 635



  \bibitem[Loore(1995)]{Loore1995} De Loore, C. \& Vanbeveren, D., 1995, A\&A, 304, 220
	
	  \bibitem[Loore(1995)]{Loore1995} De Loore, C. \& Vanbeveren, D., 1995, A\&A, 304, 220

  \bibitem[Marco(2017)]{Marco2017} De Marco, O. \& Izzard, R.G., 2017, PASA, 34, 1

	\bibitem[Mink(2013)]{Mink2013} De Mink, S.E.,  \& Langer, N., Izzard, R.G., Sana, H. \& De Koter, A., 1995, A\&A, 304, 220
      
 \bibitem[Dervis(2010)]{Dervis2010} Dervisoglu, A., Tout, C.A.\& Ibanoglu, C., 2010, MNRAS, 406, 1071

 \bibitem[Eldridge(2010)]{Eldridge2010} Eldridge, J.J., 2009, MNRAS, 400, 20

 \bibitem[Grunhut(2013)]{Grunhut2013} Grunhut, J.H., Wade, G.A., Leutenegger, M., et al., 2013, MNRAS, 428, 1686

\bibitem[Lamon(1996)]{Lamon1996} Lamontagne, R., Moffat, A.F.J., Drissen, L. et al., 1996, AJ, 112, 2227

\bibitem[Lan(2012)]{Lan2012} Langer, N., 2012, ARA\&A 50, 107
 
 \bibitem[Lan(2014)]{Lan2014} Langer, N., 2014, in Magnetic Fields throughout Stellar Evolution, IAUS302, 1

  \bibitem[Linderl(2008)]{Linder2008} Linder, N., Rauw, G., Martins, F., et al., 2008, A\&A, 489, 713
  
	\bibitem[Mahy(2011)]{Mahy2011} Mahy, L., Gosset, E., Baudin, F., et al., 2011, A\&A, 525, 101
  
  \bibitem[March(1994)]{March1994} Marchenko, S.V., Moffat, A.F.J. \& Koenigsberger, G., 1994, ApJ, 422, 810
  
	\bibitem[Massey(2000)]{Massey2000} Massey, P., Waterhouse, E., DeGioia-Eastwood, K., 2000, AJ, 119, 2214
	
  \bibitem[Menn(2014)]{Menn2014} Mennekens, N. \& Vanbeveren, D.,  2014, A\&A, 564, 134
  
    \bibitem[Meynet(2015)]{Meynet2015} Meynet, G., Chomienne, V., Ekstrom, S., et al., 2015, A\&A, 575, A60
  
  \bibitem[Naze(2017)]{Naze2017} Naze, Y., Neiner, C., Grunhut, J., et al., 2017, MNRAS (in press)
  
	\bibitem[Packet(1981)]{Packet1981} Packet, W., 1981, A\&A, 102, 17
  
  \bibitem[Petrovic(2005a)]{Petrovic2005a} Petrovic, J., Langer, N. \& van der Hucht, K.A., 2005a, A\&A, 435, 1013
  
	 \bibitem[Petrovic(2005b)]{Petrovic2005b} Petrovic, J., Langer, N., Yoon, S.-C., Heger, A., 2005b, A\&A, 435, 24

  \bibitem[Sana(2012)]{Sana2012} Sana, H., de Mink, S.E., de Koter, A., et al., 2012, Science, 337, 444
	
  \bibitem[Shara(2017)]{Shara2017} Shara, M.M., Crawford, S.M., Vanbeveren, D., et al., 2017, MNRAS, 464, 2066
    
  \bibitem[Shenar(2016)]{Shenar2016} Shenar, T., Hainich, R., Todt, H., et al., 2016, A\&A, 591, A22
  
   \bibitem[Spruit(2002)]{Spruit2002} Spruit, H.C., 2002, A\&A, 381, 923
  
  \bibitem[Tayler(1973)]{Tayler1973} Tayler, R.J., 1973, MNRAS, 161, 365
       
  \bibitem[Van den Heuvel(1993)]{VandenHeuvel1993} Van den Heuvel, E. 1993, in Saas-Fee Advanced Course, 22, eds. Nussbaumer, H. \& Orr, A., Springer-Verlag, Heidelberg, 263
  
   \bibitem[Vanbeveren(1980)]{Vanbeveren1980} Vanbeveren, D. \& Conti, P., 1980, A\&A, 88, 230
	
   \bibitem[Vanbeveren(1998a)]{Vanbeveren1998a} Vanbeveren, D., De Donder, E., Van Bever, J. et al., 1998a, NewA, 3, 443
	
	\bibitem[Vanbeveren(1998b)]{Vanbeveren1998b} Vanbeveren, D., Van Rensbergen, W. \& De Loore, C., 1998b, A\&A Review, 9, 63
	
	\bibitem[Vanbeveren(1998c)]{Vanbeveren1998c} Vanbeveren, D., Van Rensbergen, W. \& De Loore, C., 1998c, in The Brightest Binaries, Kluwer Academic Publishers, Dordrecht
	
	\bibitem[Vanbeveren(2007)]{Vanbeveren2007} Vanbeveren, D., Van Bever, J. \& Belkus, H., 2007, ApJ, 662, 107
	
	\bibitem[vdHucht(2001)]{vdHucht2001} Van der Hucht, K.A., 2001, New A. Rev., 45, 135
	
	\bibitem[Villar(2005)]{Villar2005} Villar-Sbaffi, A., St-Louis, N., Moffat, A.F.J. \& Piirola, V., 2005, ApJ, 623, 1092
	
	\bibitem[Villar(2006)]{Villar2006} Villar-Sbaffi, A., St-Louis, N., Moffat, A.F.J. \& Piirola, V., 2006, ApJ, 640, 995

  \bibitem[Zahn(1977)]{Zahn1977} Zahn, J.-P., 1977, A\&A, 57, 383 


\end{thebibliography}
\end{document}